\pdfoutput=1 
\documentclass[prl,showpacs,footinbib,twocolumn,final,superscriptaddress]{revtex4}

\usepackage{amsfonts,amsmath,amssymb,ifpdf,epsfig,graphicx}
\usepackage[usenames,dvipsnames]{color}

\usepackage{amsmath,amssymb}

\renewcommand{\leq}{\leqslant}
\renewcommand{\geq}{\geqslant}

\def\eqdef{\stackrel{\mbox{\tiny def}}{=}}     
\newcommand{\ket}[1]{|\kern.3ex#1\kern.3ex\rangle}
\newcommand{\bra}[1]{\langle\kern.3ex #1 \kern.3ex|}

\newcommand{\mean}[1]{\left\langle #1\right\rangle}
\newcommand{\smean}[1]{\langle #1\rangle}
\newcommand{\EXP}[1]{e^{#1}}         

\newcommand{\tr}[1]{\mathop{\mathrm{Tr}}\nolimits\left\{ #1 \right\}}  

\def\I{{\rm i}}

\newcommand{\derivp}[2]{\frac{\partial #1}{\partial #2}}
\newcommand{\derivf}[2]{\frac{\delta #1}{\delta #2}}
\def\D{{\rm d}}                  

\def\intpp{\smallsetminus\hspace{-0.36cm}\int}

\def\Wt{\tau_\mathrm{W}}
\def\Ht{\tau_\mathrm{H}}
\def\Nc{N}
\def\Sm{\mathcal{S}}
\def\MuZero{\mu_0}
\def\MuTwo{\mu_1}
\def\PNtau{P_\Nc(\tau)} 

\begin{document}

\title{Wigner time-delay distribution in chaotic cavities and freezing transition}

\author{Christophe Texier}
\affiliation{Universit\'e Paris Sud, CNRS, LPTMS, UMR 8626, B\^at. 100, Orsay F-91405, France}
\affiliation{Universit\'e Paris Sud, CNRS, LPS, UMR 8502, B\^at. 510, Orsay F-91405, France}
\author{Satya N. Majumdar}
\affiliation{Universit\'e Paris Sud, CNRS, LPTMS, UMR 8626, B\^at. 100, Orsay F-91405, France}

\date{March 12, 2014}


\begin{abstract}
  Using the joint distribution for proper time-delays of a chaotic cavity derived by
  Brouwer, Frahm \& Beenakker [Phys. Rev. Lett. {\bf 78}, 4737 (1997)], 
  we obtain, in the limit of large number of channels $\Nc$, the large deviation function for the 
distribution of the Wigner time-delay (the sum of proper times) by a Coulomb gas method. 
  We show that the existence of a power law tail originates from narrow resonance contributions, 
related  to a (second order) freezing transition in the Coulomb gas. 
\end{abstract}

\maketitle

The study of scattering theory in chaotic or disordered quantum systems within the random matrix theory (RMT) has been a subject of intense research for many years. Though originated in nuclear physics (see the review~\cite{GuhMulWei98}), it has major implications in condensed matter theory where it can be used to model electronic transport in mesoscopic (coherent) conductors \cite{Bee97,MelKum04}. 
The dynamics of an electron of energy $E$ is described through the $\Nc\times\Nc$ on-shell scattering matrix 
$\Sm(E)$, where $\Nc$ is the number of scattering channels.
A useful concept that characterises the temporal aspects of the scattering process is time-delay~\cite{Eis48,Wig55} undergone by an incident wave packet. This is captured by the Wigner-Smith time-delay matrix~\cite{Smi60}, $Q(E)\eqdef-\I\,\Sm(E)^\dagger\,\derivp{\Sm(E)}{E}$ (with $\hbar=1$), whose eigenvalues are the {\it proper} time-delays $\tau_1,\cdots,\tau_\Nc$.

If the system is characterised by some complex dynamics, due to the presence of disorder or chaos, 
the statistical properties of time-delays exhibit interesting universal characteristics~:
the universality of the time-delay distribution for 1D-disordered quantum mechanics was demonstrated 
in \cite{TexCom99} (see also \cite{JayVijKum89,Hei90,FarTsa94,ComTex97,OssKotGei00}, and \cite{OssFyo05} for 2D \& 3D cases).
The situation where the dynamics is chaotic has been extensively studied within RMT:
the marginal law of \textit{partial} time-delays~\cite{footnote1},
$\tilde{p}_\Nc(\tau)=\frac{1}{\Nc}\sum_a\mean{\delta(\tau-\tilde\tau_a)}$, was obtained for Gaussian unitary ensemble (GUE) of RMT indexed by $\beta=2$ \cite{FyoSom96a,FyoSom97}.
In \cite{GopMelBut96}, the time-delay distribution was derived in the $\Nc=1$ case with $\beta\in\{1,\,2,\,4\}$, corresponding to orthogonal, unitary and symplectic symmetry classes.
Using the ``alternative RMT'' introduced in \cite{BroBut97}, Brouwer and coworkers succeeded in finding the joint distribution of the inverse proper time-delays 
$\gamma_k\equiv 1/\tau_k$ (in the absence of tunable barriers at the contacts)~\cite{BroFraBee97,BroFraBee99}~:
  \begin{equation}
    \label{eq:Brouwer1997}
    P(\gamma_1,\cdots,\gamma_N)
    \propto
    \prod_{i<j}|\gamma_i-\gamma_j|^\beta
    \prod_k \gamma_k^{\beta N/2}\EXP{-\frac{\beta}{2}\gamma_k}
  \end{equation}
(the times are measured in units of the Heisenberg time $\Ht=2\pi\hbar/\Delta$, where $\Delta$ 
is the mean level spacing). 
This measure, known as the Laguerre ensemble of random matrices, also corresponds to the distribution of the (positive) 
eigenvalues of Wishart matrices $X^\dagger X$, where the matrix $X$ has size $\Nc\times(2\Nc-1+2/\beta)$ with i.i.d. Gaussian matrix 
elements.

In this article we are interested in the Wigner time delay, defined as the sum of proper (or partial) \cite{footnote1} time delays
$ \Wt \eqdef \frac{1}{\Nc}\tr{Q} = \frac{1}{N}\sum_{a=1}^\Nc\tau_a $.
This quantity is of great interest due to its close relation to the density of states (DoS) of the \textit{open system}, 
through the Krein-Friedel relation 
\cite{footnote2}~: 
$\nu(E)=\frac{1}{2\pi}\, \tr{Q(E)}=\frac{1}{2\pi}\,\Nc\Wt$.
The Wigner time delay (or related quantities such as injectance or emittance) is a 
central concept for studying charging effects, e.g. for mesoscopic capacitances \cite{GopMelBut96,ButPol05}.

\begin{figure}[!ht]
  \centering
  \includegraphics[width=0.4\textwidth]{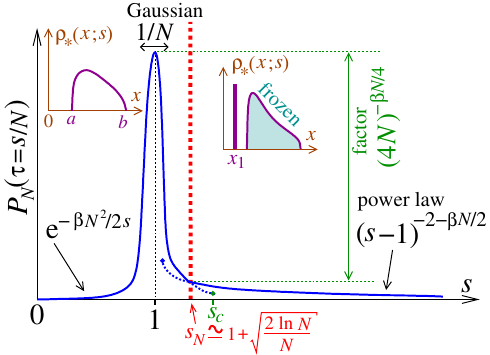}
  \caption{(color online).  
  {\it  Sketch of the distribution of $s=N\Wt$. The dashed line at $s=s_N\simeq1+(2\ln N/N)^{1/2}$ separates the two phases of the Coulomb gas with densities represented in the small figures on the left and the right respectively. }
  }
  \label{fig:PhiPlusDeS}
\end{figure}

We denote by 
$\PNtau\eqdef\smean{\delta(\tau-\frac{1}{\Nc}\sum_a\tau_a)}$ 
the Wigner time-delay distribution. Despite the fact that the joint distribution of proper times is known already for 15 years, 
little is known about the distribution of $\Wt$ for general $\Nc$:
it has been computed explicitly only for $\Nc=1$,
$
P_1(\tau)
=\frac{(\beta/2)^{\beta/2}}{\Gamma(\beta/2)}\tau^{-2-\frac{\beta}{2}}\,
\EXP{-\frac{\beta}{2\tau}}
$
\cite{GopMelBut96}
and $\Nc=2$,
$
P_2(\tau) =
\frac{ \beta^{3\beta+2} \Gamma(3(\beta+1)/2) }{ \Gamma(\beta+1)\Gamma(3\beta+2) }
\tau^{-3(\beta+1)}\,U\left( \frac{\beta+1}2 , 2(\beta+1) ; \beta/\tau \right) \,
\EXP{-\beta/\tau}
$
\cite{SavFyoSom01}, where $U(a,b;z)$ is the confluent hypergeometric function.
The distribution was conjectured to have a power law tail for large $\tau$,
$\PNtau\sim\tau^{-2-\frac{\beta}{2}\Nc}$ in \cite{FyoSom97} (for $\beta=2$) 
by using the resonance picture allowing to identify the tails of $\PNtau$ and $\tilde{p}_\Nc(\tau)$ (for a heuristic 
argument using the relation to resonance width, cf. the review \cite{Kot05}).
More recently, the first three cumulants of $\Wt$ were derived by a generating function method \cite{MezSim12}.
However a full understanding of its distribution for general $\Nc$ is still missing so far. 

In this Letter, by analysing an underlying Coulomb gas we provide a complete description of $\PNtau$ for large $\Nc$ and show that it has a rather rich behaviour including an interesting nonanalytic point which is a consequence of a freezing transition in the Coulomb gas.
Limiting behaviours of $\PNtau$ may be summarised as follows ($\Wt$ is measured in unit of~$\Ht$)~:
\begin{align}
  \label{eq:Result1}
  \PNtau 
  &\sim 
  \tau^{-\frac34\Nc^2\beta}
  \,\EXP{-\frac{\Nc\beta}{2\tau}}
  &\mbox{ for } \tau \ll \frac{1}{\Nc} 
  \\
  \label{eq:Result2}
  &\sim \exp{-\frac{\Nc^4\beta}{8}\left(\tau-\frac1\Nc\right)^2}
  &\mbox{ for } \tau \sim \frac{1}{\Nc} 
  \\
  \label{eq:Result3}
  &\sim
  \tau^{-2-\frac{\beta\Nc}{2}}
  &\mbox{ for } \tau\gg\frac{1}{\Nc}
  \:,
\end{align}
A sketch of the distribution is given in Fig.~\ref{fig:PhiPlusDeS}.
The Gaussian form around $\tau\sim1/\Nc $ in \eqref{eq:Result2} allows one to extract the mean time-delay and its variance. 
Reinstating $\Ht$, we obtain $\smean{\Wt}=\frac{\Ht}{\Nc}$. Consequently, the mean DoS reads $\smean{\nu(E)}=\Nc\smean{\Wt}/({2\pi})=1/\Delta$ as expected. Similarly, the variance can be read off \eqref{eq:Result2}
\begin{equation}
  \label{eq:VarianceWt}
  \mathrm{Var}(\Wt) \simeq \frac{4\Ht^2}{\beta N^4}
  \hspace{0.25cm}\mbox{ i.e. }\hspace{0.25cm}
  \mathrm{Var}(\nu(E))\simeq \frac4{\beta N^2\Delta^2}
  \:.
\end{equation}
Eq.~\eqref{eq:VarianceWt} was first obtained in \cite{LehSavSokSom95} for $\beta=1$. It 
agrees with the leading order of the result obtained in Ref.~\cite{MezSim12} 
$\mathrm{Var}(\Wt)=\frac{4\Ht^2}{(N+1)(N\beta-2)N^2}$.
Note also that \eqref{eq:Result3} coincides with the power law tail conjectured by Fyodorov and Sommers~\cite{FyoSom97}, 
$\PNtau\sim\tau^{-2-\frac{\beta}{2}\Nc}$.

\vspace{0.125cm}

\noindent{\it Coulomb gas.--} To derive our main results (\ref{eq:Result1},\ref{eq:Result2},\ref{eq:Result4},\ref{eq:Result3}),
we use the Coulomb gas method, originally introduced by Dyson \cite{Dys62a}.
Recently, this method 
has been suitably adopted and successfully used in a number of different contexts~: e.g. the distribution of the conductance of 
chaotic cavities \cite{VivMajBoh08,VivMajBoh10,DamMajTriViv11}, or the quantum entanglement in a random bipartite 
state \cite{FacMarParPasSca08,NadMajVer10,NadMajVer11}.
Our starting point is to rewrite the joint distribution \eqref{eq:Brouwer1997} 
of the rescaled rates $x_i=\gamma_i/\Nc$ as a Gibbs measure, 
$P(\gamma_1,\cdots,\gamma_\Nc)\propto\exp\{-\frac12\beta\Nc^2\mathcal{E}[\rho]\}$, 
with the ``energy" ${\mathcal{E}[\rho]}$ expressed as a functional of the density of the rescaled rates $\rho(x)=\frac{1}{\Nc}\sum_{i=1}^\Nc\delta(x-x_i)$.
The energy reads
\begin{align}
  \label{eq:E}
  \mathcal{E}[\rho]
  =& \int_0^\infty\D x\,(x-\ln x)\,\rho(x)
  \nonumber\\
  &-\int_0^\infty\D x\D x'\,\rho(x)\,\rho(x')\,\ln|x-x'|
\end{align}
The rescaled time-delay is $s=\Nc\Wt=\sum_i\gamma_i^{-1}$ (i.e. the DoS of the cavity in appropriate units $s=\nu(E)\Delta$).
In the limit $\Nc\to\infty$, the density $\rho(x)$ may be treated as continuous and the distribution $P_\Nc(\tau=s/\Nc)$ can be derived via a saddle point method. The 
optimal (saddle point) distribution minimizes \eqref{eq:E} with two constraints:
normalisation $\int\D x\,\rho(x)=1$ and $\int\frac{\D x}{x}\,\rho(x)=s$. This requires to minimize 
the ``free energy'' 
$\mathcal{F}[\rho]=\mathcal{E}[\rho]+\MuZero\big(\int\D x\,\rho(x)-
1\big)+\MuTwo\big(\int\frac{\D x}{x}\,\rho(x)-s\big)$, where $\MuZero$ and $\MuTwo$ are 
two Lagrange multipliers that enforce the two constraints (we neglect the subdominant contribution of entropy~\cite{DeaMaj08}).
Setting the functional derivative $\derivf{\mathcal{F}}{\rho(x_0)}=0$ gives
\begin{align}
  \label{eq:Col1}
  \hspace{-0.25cm}
   \MuZero + x_0 - \ln x_0 + \frac{\MuTwo}{x_0}
  - 2\int_a^b\D x\,\rho(x)\,\ln|x-x_0|=0
  \:.
\end{align}
where we assume that the optimal density has support over the interval $x_0\in [a,b]$.
Deriving once more with respect to $x_0$ gives
\begin{equation}
  \label{eq:Col2}
  \frac{1}{2}\left( 1 - \frac{1}{x_0} -\frac{\MuTwo}{x_0^2}\right)
  =\intpp_a^b\D x\,\frac{\rho(x)}{x_0-x}
  \:,
\end{equation}
where $\smallsetminus\hspace{-0.3cm}\int$ represents the principal part. 
This equation expresses the force balance at equilibrium, for any charge at $x_0\in [a,b]$, between 
the confining force $-V'_\mathrm{eff}(x)$ coming from the effective
potential $V_\mathrm{eff}(x)=x-\ln{x}+\frac{\MuTwo}{x}$ and the Coulomb repulsion force from other charges.  
We denote by $\rho_*(x;s)$ the solution of (\ref{eq:Col2}). The time-delay distribution then takes the scaling form 
\begin{equation}
    \PNtau \underset{N\to\infty}{\sim}
    \exp\left\{-\frac12\beta\Nc^2\Phi_-(\Nc\tau)\right\}
    \:,
\end{equation}
where the large deviation function is 
$
  \Phi_-(s)=\mathcal{E}[\rho_*(x;s)]-\mathcal{E}[\rho_*(x;1)]
$
(note that when the two constraints are fulfilled, $\mathcal{F}[\rho_*]=\mathcal{E}[\rho_*]$).
The term $\mathcal{E}[\rho_*(x;1)]$ emerges from the normalisation of \eqref{eq:Brouwer1997}, obtained by solving the same equation in 
the absence of the second constraint, i.e. for $\MuTwo=0$, which we will show to coincide with $s=1$.
Using \eqref{eq:Col1}, we may rewrite the energy of the optimal distribution as
\begin{align}
  \label{eq:Eoptimal}
  \mathcal{E}[\rho_*(x;s)] 
  =&
  \frac{\MuTwo}{2}\left(\frac1{x_0} - s\right)
  +
  \int_a^b\D x\,
  \rho_*(x;s)\,
  \\\nonumber
  &\times\left[
    \frac{x-\ln x+x_0-\ln x_0}{2}-\ln|x-x_0|
  \right]
  \:.
\end{align}

\vspace{0.125cm}

\noindent{\it Optimal distribution.--}
The integral equation \eqref{eq:Col2} may be solved thanks to a theorem due to Tricomi \cite{Tri57}. We find the optimal distribution
\begin{equation}
  \label{eq:OptimalDistribution}
    \rho_*(x;s) = \frac1{2\pi}\frac{x+c}{x^2}\sqrt{(x-a)(b-x)}
  \:,
\end{equation}
where the three parameters $a$, $b$ and $c=\MuTwo/\sqrt{ab}$ can be found by solving the three algebraic equations 
obtained by imposing the vanishing of the density at the two boundaries and the condition $\int_a^b\frac{\D x}{x}\,\rho_*(x;s)=s$. 
These equations are conveniently written in terms of the variables $v=\sqrt{ab}$ and $u=\sqrt{a/b}$. A few steps of algebra shows that $u$ solves
\begin{equation}
  \label{eq:ExpressionSigma}
  \hspace{-0.05cm}
  s=\sigma(u) \eqdef (1-u)^2
  \frac{(-u^4+16u^3+2u^2+16u-1)}{16u^2(3u^2-2u+3)} 
  \:.
\end{equation}
Then $v$, $\MuTwo$ and $c$ are given by 
$
    v = 2u \,\frac{3u^2-2u+3}{(1-u^2)^2} 
$, 
$
  \label{eq:Eq6}
    \MuTwo = -4 u^2 \frac{(u^2-6u+1)(3u^2-2u+3)}{(1-u^2)^4}
$
and
$c=\frac{\MuTwo}{v}=-2u\,\frac{(u^2-6u+1)}{(1-u^2)^2}$.

\vspace{0.125cm}

\begin{figure}[!ht]
  \centering
  \includegraphics[width=0.45\textwidth]{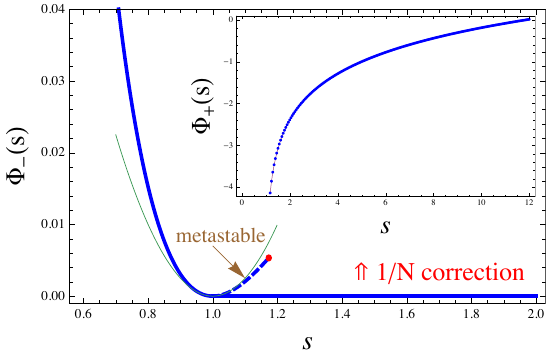}
  \caption{(color online).
  {\it Large deviation function $\Phi_-(s)$ (i.e. rescaled energy of the gas). The freezing transition takes place at $s_\infty=1$. The metastable branch terminates at $s_c=1.1738...$
  Inset~: Large deviation function $\Phi_+(s)$ [i.e. $1/N$ correction to the rescaled energy]. }
  }
  \label{fig:PhideS}
\end{figure}

\noindent{\it Most probable values.--}
We first analyse the distribution $\PNtau$ in the vicinity of its maximum. $\mathcal{E}[\rho_*(x;s)]$ is minimised, i.e. $\PNtau$ is maximized, by removing the constraint $\int_a^b\frac{\D x}{x}\,\rho(x)=s$, i.e. by setting $\MuTwo=0$. For convenience we introduce the roots $x_\pm=3\pm2\sqrt2$ of the polynomial $u^2-6u+1$. For $\MuTwo=0$, Eq.~(\ref{eq:ExpressionSigma}) has solution 
$u=\sqrt{x_-/x_+}=x_-$ with $v=1$ and $s=1$ and consequently $a=x_-=0.171...$ and $b=x_+=5.828...$
 In this case we 
recover the Mar\v{c}enko-Pastur (MP) law \cite{MarPas67}
\begin{equation}
 \rho_*(x;1)=\frac1{2\pi x}\sqrt{(x-x_-)(x_+-x)}
\end{equation} 
Expansion of Eq.~(\ref{eq:ExpressionSigma}) 
around the MP point leads to
$s-1\simeq-\frac{x_+}{\sqrt2}(u-x_-)$, hence
$v\simeq1+\frac{3x_+}{2\sqrt2}(u-x_-)\simeq1-\frac32(s-1)$
and $c\simeq\MuTwo\simeq-\frac12(s-1)$.
The corresponding energy \eqref{eq:Eoptimal} may be conveniently obtained by choosing $x_0=1$~: we see that the first term is quadratic $\frac14(s-1)^2$~; we check numerically that the remaining integral term is constant, equal to $\mathcal{E}[\rho_*(x;1)]=3-2\ln2$, up to higher order corrections [numerical fit gives a correction $\frac14(s-1)^3$]. Therefore we conclude that 
$
   \Phi_-(s) \underset{s\sim1}{\simeq} \frac14(s-1)^2
$, 
i.e. Eq.~\eqref{eq:Result2} (the parabolic behaviour is compared to the numerical calculation of the integral \eqref{eq:Eoptimal} in Fig.~\ref{fig:PhideS}).

\begin{figure}[!ht]
  \centering
  \includegraphics[width=0.45\textwidth]{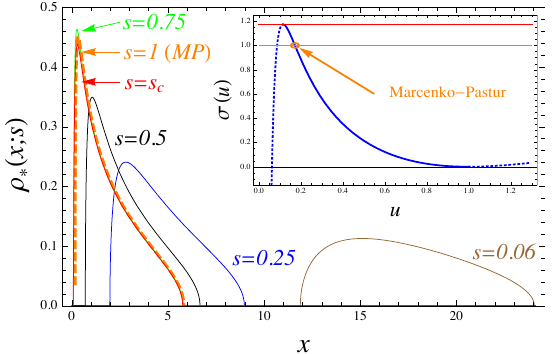}
  \caption{(color online) 
  {\it The optimal density of eigenvalues for different values of $s$~; when $s$ increases, the density eventually freezes to the MP law (dashed line).}
  }
  \label{fig:RhoStar}
\end{figure}

\vspace{0.125cm}

\noindent{\it Large deviations for $s\to0$.--}
Expansion of \eqref{eq:ExpressionSigma} for $s\to0$ gives $u=1-\sqrt{2s}+s+\mathcal{O}(s^{3/2})$, hence
$v=\frac{1}{s}+\mathcal{O}(s^0)$.
The support of the distribution is given by 
$a=\frac{1}{s}\big[1-\sqrt{2s}+\mathcal{O}(s)\big]$ and $b=\frac{1}{s}\big[1+\sqrt{2s}+\mathcal{O}(s)\big]$
(the Lagrange multiplier is $\MuTwo=\frac{1}{s^2}+\mathcal{O}(s^{-1})$).
The optimal distribution ressembles the semi-circle law centered around~$1/s$~:
\begin{equation}
  \label{eq:LDL}
  \rho_*(x;s) \underset{s\to0}{\simeq}
  \frac{1}{\pi} \,
  \sqrt{2s-\left(s\,x-1\right)^2}
  \:.
\end{equation}
This was expected~: when $s\to0$, the eigenvalues $\{x_i\}$ of the Wishart matrix are constrained to be very large and they do not feel the spectrum boundary at $x=0$. Hence, their distribution coincides with the Wigner semi-circle law for the usual Gaussian ensembles of random matrices.
The energy may be conveniently calculated by choosing $x_0=1/s$~; this makes the first term of \eqref{eq:Eoptimal} vanish. The leading order of the integral term is straightforwardly calculated from \eqref{eq:LDL}~: we deduce 
$
  \Phi_-(s)\underset{s\to0}{\simeq}\frac{1}{s}+\frac32\ln s-\frac52(1-\ln2)
$, 
thus proving~\eqref{eq:Result1}. 
The factor $\exp{-\frac{\Nc\beta}{2\tau}}$ is in perfect agreement with
the exact results for $\Nc=1$ \& $2$ mentioned earlier.

\vspace{0.125cm}

\noindent{\it Large deviations for $s\geq1$ -- Freezing transition.--}
As $s$ increases, it eventually reaches a finite value corresponding to the maximum of the function $\sigma(u)$ (inset of Fig.~\ref{fig:RhoStar}), at $u_c=\frac13\big[ 1+2 (2^{1/3}-2^{2/3}) \big]=0.115...$ giving $s_c=\sigma(u_c)=\frac{10+6\times2^{1/3}-11\times2^{2/3} }{3\,(6-6\times2^{1/3}+2^{2/3})}=1.1738...$. Then $a=-c$, which leads to a somewhat unusual form
\begin{equation}
  \label{eq:CriticalDistrib}
  \rho_*(x;s_c) = \frac{1}{2\pi x^2}(x-a)^{3/2}(b-x)^{1/2}
  \:.
\end{equation}
For $s>s_c$, \eqref{eq:ExpressionSigma} has no longer physical (real) solutions. In this case, the saddle
point turns out to have a different solution where a single isolated charge,
say at $x_1$, splits off the main body of the density and carries a macroscopic weight (see Fig.~\ref{fig:PhiPlusDeS}). 
A similar scenario occurs in the study of quantum entanglement in random bipartite state~\cite{FacMarParPasSca08,NadMajVer10,NadMajVer11}.
We decompose the density as  
$
\rho(x) = \frac{1}{\Nc}\delta(x-x_1)
  + \tilde{\rho}(x)
$
where  
$\tilde{\rho}(x)=\frac{1}{\Nc}\sum_{i>1}\delta(x-x_i)$
is still treated as a continuous density.
The energy 
\begin{equation}
  \label{eq:EnergyAboveSc}
  \mathcal{E}[\rho] = \mathcal{E}[\tilde\rho]
  +\frac{x_1-\ln x_1}{\Nc}
  -\frac{2}{\Nc}\int\D x\, \tilde{\rho}(x) \, \ln(x-x_1)
\end{equation}
must be minimized under the two constraints $\int\D x\,\tilde\rho(x)=1-\frac1N$ and $\int\D{x}\,\frac{\tilde\rho(x)}{x}=s-\frac1{Nx_1}$.
This leads to the two equilibrium conditions
\begin{eqnarray}
  \label{eq:Col1overN1}
  \hspace{-0.5cm}
  \frac{1}{2}\left( 1 - \frac{1}{x_0} -\frac{\MuTwo}{x_0^2}\right) - \frac{1}{\Nc }\frac{1}{x_0-x_1}
  &=&\intpp_a^b\D x'\,\frac{\tilde\rho(x')}{x_0-x'}
  \\
  \label{eq:Col1overN2}
  \hspace{-0.5cm}
  \frac{1}{2}\left( 1 - \frac{1}{x_1} -\frac{\MuTwo}{x_1^2}\right)
  &=&\int_a^b\D x'\,\frac{\tilde\rho(x')}{x_1-x'}
  \:,
\end{eqnarray}
$\forall\:x_0\in[a,b]$ and $x_1<a$. 
We show that a consistent picture is the freezing of the density $\tilde\rho(x)$ while the isolated charge goes to zero $x_1\to0$. When $\Nc\to\infty$, the r.h.s. of \eqref{eq:Col1overN2} reaches a constant value as $x_1\to0$~; so does the l.h.s. iff $\MuTwo\simeq-x_1\to0^-$. Hence the solution of \eqref{eq:Col1overN1} is the MP law~: $\tilde\rho_*(x;s)=\rho_*(x;1)+\mathcal{O}(\Nc ^{-1})$. The rescaled time delay splits into the contribution of the isolated charge and of $\tilde\rho$ as $s=\frac{1}{\Nc x_1}+1$, i.e. $x_1=1/[\Nc(s-1)]$.
In fact this analysis holds for any $s>1$ (and not only $s\geq s_c$)~: 
the energy \eqref{eq:EnergyAboveSc} of this new phase coincides with the energy of the MP solution, 
up to $1/\Nc$ corrections. Therefore for $1<s\leq s_c$ we have found another phase with a lower energy, which shows that 
the branch obtained previously (with compact solution \eqref{eq:OptimalDistribution} over $[a,b]$ for $s<s_c$ as well as \eqref{eq:CriticalDistrib} for $s=s_c$) is actually \textit{metastable} (Fig.~\ref{fig:PhideS}). 
In the (thermodynamic) limit $\Nc\to\infty$, the energy of the gas vanishes for all $s>1$,
while for $s<1$, it behaves as $\frac{1}{4}\, (1-s)^2$ as mentioned earlier (Fig.~\ref{fig:PhideS}). This then results in a \textit{second order} phase transition at $s=1$. We call
this a \textit{freezing} transition, because for $s>1$, energy freezes to the value $0$
in the thermodynamic limit and also the bulk density freezes to the MP distribution.

One can analyse more precisely this new \textit{frozen} phase by computing the $1/\Nc$ corrections to the energy.
For large enough $s$, Eq.~\eqref{eq:EnergyAboveSc} is dominated by the logarithmic term $-\frac1\Nc\ln x_1$, i.e. $\mathcal{E}[\rho_*(x;s)]\simeq(\cdots)+\frac{1}{\Nc}\ln\big[\Nc(s-1)\big]$.
We get the power law tail $\PNtau\sim (s-1)^{-\tilde\theta-\frac{\beta}{2}\Nc}$, where $\tilde\theta$ is some exponent of order $\Nc^0$ introduced in order to account for $\Nc^{-2}$ corrections to $\mathcal{E}[\rho]$.
This exponent may be determined as follows~: 
when $\Wt>1/\Nc$, most of the proper times are described by the frozen density (the MP law), i.e. $\tau_i\in[x_-/\Nc,x_+/\Nc]$ for $i>1$ with $\sum_{i>1}1/\tau_i=1$, while one proper time becomes much larger and carries a ``macroscopic'' contribution, $\tau_1=s-1=\Nc\Wt-1$. In the scattering problem, this is interpreted as the large contribution of a narrow \textit{resonance}.
Writing 
$\PNtau=\int\D\gamma_1\cdots\D\gamma_\Nc\,\delta(\Nc\tau-1/\gamma_1-1)\,P(\gamma_1,\cdots,\gamma_\Nc)$ and using \eqref{eq:Brouwer1997} leads to $\tilde\theta=2$, hence Eq.~\eqref{eq:Result3}.

A more precise analysis of Eqs.~(\ref{eq:Col1overN1},\ref{eq:Col1overN2}) leads to introduce the large deviation function 
$
  \Phi_+(s) = 
  \Nc\, \big(
    \mathcal{E}[\rho_*(x;s)] - \mathcal{E}[\rho_*(x;1)] 
  \big)
  - \ln\Nc
$
giving the scaling form
\begin{equation}
  \label{eq:ScalingPNdeSAbove}
  \hspace{-0.15cm}
  \PNtau \sim \Nc^{-\frac{\beta\Nc}{2}}
  \exp\left\{-\frac{\beta\Nc}{2}\Phi_+(\Nc\tau)\right\}
  \hspace{0.25cm}\mbox{for }\tau>\frac{s_\Nc}{\Nc}
\end{equation}
One obtains that~\cite{GraMajTex14}
 $\Phi_+(s)=\ln(s-1)-1-2\ln2$ 
(c.f. inset of Fig.~\ref{fig:PhideS}). 
The logarithmic behaviour to the power law tail \eqref{eq:Result3}.
For finite $\Nc$, the energy functions characterising the two phases cross for $s=s_\Nc$ such that $\Phi_-(s_\Nc)=\frac{1}{\Nc}\big[\Phi_+(s_\Nc)+\ln\Nc\big]$. 
Using the limiting behaviours for $s\to1$, we obtain the finite $\Nc$ correction to the position of the phase transition~: 
$s_\Nc\simeq1+(2\ln\Nc/\Nc)^{1/2}$.

\vspace{0.125cm}

\noindent{\it Conclusion.--}
In summary, by using a Coulomb gas approach, we have analysed the large deviation functions controlling the Wigner time-delay distribution in the limit of large number of conducting channels. We have shown that the distribution exhibits a rich structure. In particular, its power law tail is related to a freezing transition in the Coulomb gas, corresponding to large contributions to $\Wt$ of resonant states in the original scattering problem.
We have also performed a Monte-Carlo simulation of the Coulomb gas up to 1600 charges and found good agreement with our analytical results (details will be published elsewhere).

Several questions remain open~: 
({\it i}) a more precise treatment of $1/\Nc$ corrections would be desirable.
({\it ii}) 
The role of tunneling couplings at the contacts and the crossover between GOE and GUE symmetries were studied in \cite{FyoSavSom97} for the marginal law $\tilde{p}_\Nc(\tau)$. Similar questions naturally arise for the Wigner time delay distribution and might be relevant for experimental purposes.
({\it iii}) 
The starting point of our calculation, Eq.~\eqref{eq:Brouwer1997}, describes the usual random matrix ensembles~;
the distribution of $\Wt$ was also obtained in \cite{FyoOss04} for a chiral-GUE ensemble when $\Nc=1$.
Extension of our analysis to such cases  would be certainly interesting, 
in particular with the growing interest in the study of new symmetry classes of disordered systems.

\vspace{0.125cm}

\noindent{\it Acknowledgments.--}
C.T. acknowledges stimulating discussions with D.~Savin, N.~Simm \& D. Villamaina. S.N.M acknowledges
ANR grant 2011-BS04-013-01 WALKMAT and support from the Indo-French Centre for the Promotion of Advanced Research under Project 4604-3.

Note added after publication~: the large deviation function $\Phi_+(s)$ has been corrected thanks to a remark of Aur\'elien Grabsch. Details will be published elsewhere~\cite{GraMajTex14}.


\begin{thebibliography}{100}

\bibitem{GuhMulWei98}
T.~Guhr, A.~M{\"u}ller-Groeling, and H.~A. Weidenm\"uller,
Random-matrix theories in quantum physics: common concepts,
Phys. Rep. {\bf 299}(4/6), 189--425 (1998).

\bibitem{Bee97}
C.~W.~J. Beenakker,
Random-Matrix theory of quantum transport,
Rev. Mod. Phys. {\bf 69}(3), 731--808 (1997).

\bibitem{MelKum04}
P.~A. Mello and N.~Kumar,
{\em Quantum transport in mesoscopic systems -- Complexity and
  statistical fluctuations},
Oxford University Press, 2004.

\bibitem{Eis48}
L.~Eisenbud,
PhD thesis, Princeton, 1948.

\bibitem{Wig55}
E.~P. Wigner,
Lower limit for the energy derivative of the scattering phase shift,
Phys. Rev. {\bf 98}(1), 145--147 (1955).

\bibitem{Smi60}
F.~T. Smith,
Lifetime matrix in collision theory,
Phys. Rev. {\bf 118}(1), 349--356 (1960).

\bibitem{TexCom99}
C.~Texier and A.~Comtet,
Universality of the Wigner time delay distribution for
  one-dimensional random potentials,
Phys. Rev. Lett. {\bf 82}(21), 4220--4223 (1999).

\bibitem{JayVijKum89}
A.~M. Jayannavar, G.~V. Vijayagovindan, and N.~Kumar,
Energy dispersive backscattering of electrons from surface resonances
  of a disordered medium and $1/f$ noise,
Z. Phys. B -- Condens. Matter {\bf 75}, 77--79 (1989).

\bibitem{Hei90}
J.~Heinrichs,
Invariant embedding treatment of phase randomization and electrical
  noise at disordered surfaces,
J.~Phys.~Cond. Matter {\bf 2}, 1559--1568 (1990).

\bibitem{FarTsa94}
W.~G. Faris and W.~J. Tsay,
Time delay in random scattering,
SIAM J. Appl. Math. {\bf 54}(2), 443--455 (1994).

\bibitem{ComTex97}
A.~Comtet and C.~Texier,
On the distribution of the Wigner time delay in one-dimensional
  disordered systems,
J.~Phys.~A: Math. Gen. {\bf 30}, 8017--8025 (1997).

\bibitem{OssKotGei00}
A.~Ossipov, T.~Kottos, and T.~Geisel,
Statistical properties of phases and delay times of the
  one-dimensional Anderson model with one open channel,
Phys. Rev. B {\bf 61}, 11411--11415 (2000).

\bibitem{OssFyo05}
A.~Ossipov and Y.~V. Fyodorov,
Statistics of delay times in mesoscopic systems as a manifestation of
  eigenfunction fluctuations,
Phys. Rev.~B {\bf 71}(12), 125133 (2005).

\bibitem{footnote1}  The partial time delays $\tilde\tau_a$ are defined as derivatives of the phase shifts [phases of the eigenvalues of $\mathcal{S}(E)$]. $\tilde\tau_a$ measures the time spent in the scattering region by a wave packet in scattering channel $a$ with narrow dispersion in energy around $E$. Because derivation and diagonalisation do not commute, partial times $\tilde\tau_a$ differ from proper times $\tau_a$~; they however satisfy the sum rule $\sum_a\tilde\tau_a=\sum_a\tau_a$.

\bibitem{FyoSom96a}
Y.~V. Fyodorov and H.-J. Sommers,
Parametric correlations of scattering phase shifts and fluctuations
  of delay times in few-channel chaotic scattering,
Phys. Rev. Lett. {\bf 76}(25), 4709 (1996).

\bibitem{FyoSom97}
Y.~V. Fyodorov and H.-J. Sommers,
Statistics of resonance poles, phase shift and time delays in quantum
  chaotic scattering: Random matrix approach for systems with broken
  time-reversal invariance,
J. Math. Phys. {\bf 38}(4), 1918--1981 (1997).

\bibitem{GopMelBut96}
V.~A. Gopar, P.~A. Mello, and M.~B{\"u}ttiker,
Mesoscopic capacitors: a statistical analysis,
Phys. Rev. Lett. {\bf 77}(14), 3005 (1996).

\bibitem{BroBut97}
P.~W. Brouwer and M.~B{\"u}ttiker,
Charge-relaxation and dwell time in the fluctuating admittance of a
  chaotic cavity,
Europhys. Lett. {\bf 37}(7), 441--446 (1997).

\bibitem{BroFraBee97}
P.~W. Brouwer, K.~M. Frahm, and C.~W. Beenakker,
Quantum mechanical time-delay matrix in chaotic scattering,
Phys. Rev. Lett. {\bf 78}(25), 4737 (1997).

\bibitem{BroFraBee99}
P.~W. Brouwer, K.~M. Frahm, and C.~W. Beenakker,
Distribution of the quantum mechanical time-delay matrix for a
  chaotic cavity,
Waves Random Media {\bf 9}, 91--104 (1999).

\bibitem{footnote2}  Precisely, $\nu(E)$ is defined as the local DoS integrated inside the scattering region.  As pointed out in several papers of B\"uttiker (cf. \cite{ButPol05} for instance) $\nu(E)$ should be rather obtained by considering the derivative of the scattering matrix with respect to a uniform internal potential, instead of a derivative with respect to the energy. However the difference decays with the energy as $1/E$, i.e. faster than the DoS in any dimension (see Eq.~(53) of \cite{TexBut03}).


\bibitem{ButPol05}
M.~B\"{u}ttiker and M.~L. Polianski,
Charge fluctuation in open chaotic cavities,
J.~Phys.~A: Math. Theor. {\bf 38}, 10559--10585 (2005).

\bibitem{TexBut03}
C.~Texier and M.~B{\"u}ttiker,
Local Friedel sum rule in graphs,
Phys. Rev.~B {\bf 67}(24), 245410 (2003).

\bibitem{SavFyoSom01}
D.~V. Savin, Y.~V. Fyodorov, and H.-J. Sommers,
Reducing nonideal to ideal coupling in random matrix description of
  chaotic scattering: Application to the time-delay problem,
Phys. Rev. E {\bf 63}, 035202 (2001).

\bibitem{Kot05}
T.~Kottos,
Statistics of resonances and delay times in random media: beyond
  random matrix theory,
J.~Phys.~A: Math. Theor. {\bf 38}, 10761--10786 (2005).

\bibitem{MezSim12}
F.~Mezzadri and N.~J. Simm,
$\tau$-Function Theory of Quantum Chaotic Transport
  with $\beta=1,\,2,\,4$, Commun. Math. Phys. {\bf 324}, 465--513 (2013).

\bibitem{LehSavSokSom95}
N.~Lehmann, D.~V. Savin, V.~V. Sokolov and H.-J. Sommers, Time delay
  correlations in chaotic scattering: random matrix approach, Physica D {\bf
  86}, 575--585 (1995).

\bibitem{Dys62a}
F.~J. Dyson,
Statistical Theory of the Energy Levels of Complex Systems,
J. Math. Phys. {\bf 3}(1), 140--156 (1962)~;
\textit{ibid} {\bf 3}(1), 157--165 (1962)~;
\textit{ibid} {\bf 3}(1), 166--175 (1962).

\bibitem{VivMajBoh08}
P.~Vivo, S.~N. Majumdar, and O.~Bohigas,
Distributions of Conductance and Shot Noise and Associated Phase
  Transitions,
Phys. Rev. Lett. {\bf 101}(21), 216809 (2008).

\bibitem{VivMajBoh10}
P.~Vivo, S.~N. Majumdar, and O.~Bohigas,
Probability distributions of linear statistics in chaotic cavities
  and associated phase transitions,
Phys. Rev. B {\bf 81}, 104202 (2010).

\bibitem{DamMajTriViv11}
K.~Damle, S.~N. Majumdar, V.~Tripathi, and P.~Vivo,
Phase Transitions in the Distribution of the Andreev Conductance of
  Superconductor-Metal Junctions with Multiple Transverse Modes,
Phys. Rev. Lett. {\bf 107}, 177206 (2011).

\bibitem{FacMarParPasSca08}
P.~Facchi, U.~Marzolino, G.~Parisi, S.~Pascazio, and A.~Scardicchio,
Phase Transitions of Bipartite Entanglement,
Phys. Rev. Lett. {\bf 101}, 050502 (2008).

\bibitem{NadMajVer10}
C.~Nadal, S.~N. Majumdar, and M.~Vergassola,
Phase Transitions in the Distribution of Bipartite Entanglement of a
  Random Pure State,
Phys. Rev. Lett. {\bf 104}, 110501 (2010).

\bibitem{NadMajVer11}
C.~Nadal, S.~N. Majumdar, and M.~Vergassola,
Statistical distribution of quantum entanglement for a random
  bipartite state,
J. Stat. Phys. {\bf 142}(2), 403--438 (2011).

\bibitem{DeaMaj08}
D.~S. Dean and S.~N. Majumdar,
Extreme value statistics of eigenvalues of Gaussian random matrices,
Phys. Rev. E {\bf 77}, 041108 (2008).

\bibitem{Tri57}
F.~G. Tricomi,
{\em Integral equations},
Interscience, London, 1957,
Pure Appl. Math. V.

\bibitem{MarPas67}
V.~A. Mar{\v c}enko and L.~A. Pastur,
Distribution of eigenvalues for some sets of random matrices,
Matem. Sbornik {\bf72}(114), 507--536 (1967).

\bibitem{FyoSavSom97}
Y.~V. Fyodorov, D.~V. Savin and H.-J. Sommers, Parametric correlations of phase
  shifts and statistics of time delays in quantum chaotic scattering: Crossover
  between unitary and orthogonal symmetries, Phys. Rev. E {\bf 55},
  R4857--R4860 (1997).

\bibitem{FyoOss04}
Y.~V. Fyodorov and A.~Ossipov,
Distribution of the Local Density of States, Reflection Coefficient,
  and Wigner Delay Time in Absorbing Ergodic Systems at the Point of Chiral
  Symmetry,
Phys. Rev. Lett. {\bf 92}, 084103 (2004).


\bibitem{GraMajTex14}
A.~Grabsch, S. Majumdar and C. Texier,
unpublished (2014).


\end{thebibliography}

\end{document}